\documentstyle[epsf,aps,prb,preprint,tighten,epsfig]{revtex}
\input{epsf}
\begin {document}
\draft
\title{Scenario of pseudogap in underdoped cuprate}
\author{Ping Lou$^1$ and Hang-sheng Wu$^2$}
\address{1.Department of Physics,Anhu University ,Hefei 230039, People's Republic of China\\
2.Department of Physics, University of Science and Technology of China,
                     Hefei 230026, People's Republic of China }
\maketitle
\begin{abstract}
So far, the theories of the cuprate pseudogap may be broadly divided into two main schools of though. One is based upon the idea that view the pseudogap as deriving from some of precursor superconductivity. Another assumes that associate the pseudogap phenomena with magnetic pairing of some sort. Both the scenario of superconducting fluctuation and the scenario of magnetic pairing of some sort on the pseudogap in underdoped cuprate are based upon the idea that a fluctuation of a certain type exists at temperatures higher than the usual superconducting transition temperature and has an effect on the single particl self-energy and just such a self-energy leads to the gap-like structure of the spectral weight. In this paper we argue that although a fluctuation of a certain type has an effect on the origin of the pseudogap, it is not the crucial factor of the origin of the pseudogap in underdoped cuprate. We analysis the scenario of the pseudogap in the underdoped cuprate based on the model that can bring about the spin density wave (SDW) and find the new dispersion characteristics of the high temperature cuprate superconductors (excepting two different saddle points located at ($\pm\pi$, 0) and (0, $\pm\pi$)) and suggest that the pseudogap formation might been due to the unique interior of the underdoped cuprate.
\draft
\vskip 0.5cm
\end{abstract}

\noindent {\bf Introduction}  
\vspace{.5cm}

Recently, the various experiments have established the fact that the underdoped cuprate superconductors exhibit a pseudogap behavior bellow a characteristic temperature $T^\ast$ which can be well above the superconducting transition temperature $T_c$. The so-called ``pseudogap'' means a partial gap, i.e. some regions of the Fermi surface become gapped while other parts retain their conducting properties and with increased doping the gapped portion diminishes and the materials become more metallic$^1$. In the theoretical respect a variety of theoretical scenarios have also proposed for the origin of the pseudogap, however, no consensus has been reached so far, which of the various microscopic theories is the correct one. The theories of the cuprate pseudogap may be broadly divided into two main schools of though. One is based upon the idea that view the pseudogap as deriving from some of precursor superconductivity, for example Refs.[2-11]. Another assumes that associate the pseudogap phenomena with magnetic pairing of some sort, for example Refs.[12-18]. Both the scenario of superconducting fluctuation and the scenario of magnetic pairing of some sort on the pseudogap in underdoped cuprate are based upon the idea that a fluctuation of a certain type exists at temperatures higher than the usual superconducting transition temperature and has an effect on the single particl self-energy and just such a self-energy leads to the gap-like structure of the spectral weight. In this paper we argue that although a fluctuation of a certain type has an effect on the origin of the pseudogap, it is not the crucial factor of the origin of the pseudogap in underdoped cuprate. We analysis the scenario of the pseudogap in the underdoped cuprate based on the model that can bring about the spin density wave (SDW) and find the new dispersion characteristics of the high temperature cuprate superconductors (excepting two different saddle points located at ($\pm\pi$, 0) and (0, $\pm\pi$)) and suggest that the pseudogap formation might been due to the unique interior of the underdoped cuprate.
 
\vspace{.5cm}
\noindent {\bf Model and Calculations} 
\vspace{.5cm} 

  We consider a two-dimension square lattice with the kinetic energy given by,
 \begin{equation}
H_o= \sum_{k\sigma}(\overline{\varepsilon}_{k}-\overline{\mu})a_{k\sigma}^{\dagger}a_{k\sigma} 
\end{equation}
with
\begin{equation}
\overline{\varepsilon}_{k}=-2t(\cos k_x+\cos k_y)-4t^{\prime}\cos k_x\cos k_y
\end{equation}

where $a_{k\sigma}(a_{k\sigma}^{\dagger})$ is the annihilation (creation) operator for the electron. The $k$ sum is extended over the first Brillouin zone (Fig.1). $t$ (nearest neighbor), $t^\prime$(next nearest neighbor) and $\overline{\mu}$ (chemical potential). In this paper we consider $t>0$ and $t^{\prime}<0$ case only. It is well known that the form of the dispersion law (2) is characterized by two different saddle points locate at ($\pm\pi$, 0) and (0, $\pm\pi$) with the energy $\overline{\varepsilon}_{sd}=4t^{\prime}$, which is shown in Fig.1.

In order to further find the dispersion characteristics of the high temperature cuprate superconductors (Besides two different saddle points located at ($\pm\pi$, 0) and (0, $\pm\pi$)), we take the Brillouin zone as shown in Fig.2, which consists of the region A and B including ($\pm\pi$, 0) and (0, $\pm\pi$) respectively. Then equation (1) is changed into following form

\begin{equation}
H_o=\sum_{k\sigma}(\varepsilon_{k}-\mu)a_{k\sigma}^{\dagger}a_{k\sigma}+\sum_{k\sigma}(\varepsilon_{k+Q}-\mu)a_{k+Q\sigma}^{\dagger}a_{k+Q\sigma}
\end{equation}

where $\mu=\overline{\mu}-4t^{\prime}$, $Q=(\pi,\pi)$ and

\begin{equation}
\varepsilon_{k}=-2t(-\cos k_x+\cos k_y)+4t^{\prime}(\cos k_x\cos k_y-1)     
\end{equation}

\begin{equation}
\varepsilon_{k+Q}=-2t(\cos k_x-\cos k_y)+4t^{\prime}(\cos k_x\cos k_y-1)     
\end{equation}

   The $k$ sums in Eqs.(3) are extended only up to the reduced magnetic Brillouin zone boundary. 
On the other hand we have taken $4t^{\prime}$ as new zero point of energy. New chemical potential is defined by $\mu$ which is adjusted to obtain the required doping $\delta$. For the underpoed regime, $\mu>0$; For the overdoped regime, $\mu<0$; For the optimally doped, $\mu=0$. For example, when $t^{\prime}/t=-0.25$ we have: (i) $\mu/t=0$, $\delta=0.22$, the optimally doped; (ii) $\mu/t=0.2$, $\delta=0.1$, in the underpoed regime; (iii) $\mu/t=-0.12$, $\delta=0.3$,  in the overdoped regime. If we assumed that $\varepsilon_{k}$ and $\varepsilon_{k+Q}$ are the electron-like and the hole-like band respectively, we can give that the dispersion characteristics of the high temperature cuprate superconductors (besides the saddle points) and show in Fig.2 and Tabe-1. It is indeed that beside the saddle points there are the necklace regions (in these regions $\varepsilon_{k}>0$, while $\varepsilon_{k+Q}>0$.) which is indicated by the shadow regions in our choice of the first Brillouin zone (i.e.$A^{\prime}_+$ and $B^{\prime}_+$ ) and only appears when $t^{\prime}\not=0$ . On the other hand in Fig.3 we give the Fermi surface characteristics. For the overdoped ($\mu<0$) the Fermi surface is at the out of the necklace regions. For the underdoped ($\mu>0$) some portion of the Fermi surface is in the necklace regions, while other parts is at the out of the necklace regions and with decreased doping the portion out of the necklace increases (In Fig.3 it are represented by the thick sections.). When the system opens gap, the Fermi surface of out of the necklace regions open gap, while the Fermi surface in the necklace regions are subjected to small effect only and can remain metallic.  These just are the unique interior for the pseudogap formation in the underdoped cuprate (As for the role of incommensurations, because on the Fermi surface when the $k_F$ of $\varepsilon_{k}$ adds
$\tilde{Q}$ (where $\tilde{Q}$ =($\pi\pm\delta\pi$,$\pi$) or ($\pi$,$\pi\pm\delta\pi$ )), then $\varepsilon_{k}$ changes into $\varepsilon_{k+\tilde{Q}}$. Except in the nearby regions of the saddle points these characteristics still survive and in this context we ignore the role of incommensurations.).

Next we consider the interaction term. For the mean-field theory of the SDW, the interaction term can been written as

\begin{equation}
H_{SDW}=-\sum_{k\sigma}(\Delta(k)a_{k+Q\sigma}^{\dagger}\sigma a_{k\sigma}+h.c)
\end{equation}
where the $k$ sums is extended only up to the reduced magnetic Brillouin zone boundary (following same). $\Delta(k)$ is the SDW energy gap parameter and defined as

\begin{equation}
\Delta(k)=\frac{1}{N}\sum_{k^{\prime}\sigma}V_{kk^{\prime}}<a_{k^{\prime}+Q\sigma}^{\dagger}\sigma a_{k^{\prime}\sigma}>
\end{equation}

where $V_{kk^{\prime}}$ is $U$ for Hubbard interaction and $U(\cos k_x+\cos k_y)(\cos k^{\prime}_x+\cos k^{\prime}_y)$ for the d-wave separable potential.
   By Eqs.(3) and Eqs.(6) we can write the Hamiltonian of the model in the matrix representation as

\begin{equation}
H=\sum_{k\sigma}\Psi^{\dagger}_{k\sigma}\hat{H}_M\Psi_{k\sigma}
\end{equation}
where the Hamiltonian matrix ($\hat{H}_M$) is obtained as

\begin{equation}
\hat{H}_M=\left(\matrix{\varepsilon_{k}-\mu   & -\Delta(k)\sigma \cr
   -\Delta(k)\sigma  & \varepsilon_{k+Q}-\mu}
     \right)
\end{equation}
and 
           
\begin{equation}
\Psi^{\dagger}_{k\sigma}=\left(\matrix{a^{\dagger}_{k\sigma} & a^{\dagger}_{k+Q\sigma}}\right)
\end{equation}

By the following Bogoliubov transformation $\Psi_{k\sigma}=\hat{U}_{k\sigma}\Phi_{k\sigma}$, i.e.
\begin{equation}
\left(\matrix{a_{k\sigma} \cr a_{k+Q\sigma}}\right)=\left(\matrix{\mu_{k} & -\Delta(k)\nu_{k} \cr \Delta(k)\nu_{k} & \mu_{k}}\right)\left(\matrix{\alpha_{k\sigma} \cr \gamma_{k\sigma}}\right)
\end{equation}
where 
$\mu_{k}=\frac{1}{2}\sqrt{1+\frac{\varepsilon_{k}-\varepsilon_{k+Q}}{2E(k)}}$,  $\nu_{k}=\frac{1}{2}\sqrt{1-\frac{\varepsilon_{k}-\varepsilon_{k+Q}}{2E(k)}}$  and  $E(k)=\sqrt{(\frac{\varepsilon_{k}-\varepsilon_{k+Q}}{2})^2+|\Delta(k)|^2}$    
we can obtain the diagonalised SDW Hamiltonian as
\begin{equation}
H=\sum_{k\sigma}(\varepsilon_{1}(k)\alpha^{\dagger}_{k\sigma}\alpha_{k\sigma}+\varepsilon_{2}(k)\gamma^{\dagger}_{k\sigma}\gamma_{k\sigma})
\end{equation}
  with
\begin{equation}
\varepsilon_{1}(k)=\frac{\varepsilon_{k}+\varepsilon_{k+Q}}{2}-\mu+\sqrt{(\frac{\varepsilon_{k}-\varepsilon_{k+Q}}{2})^2+|\Delta(k)|^2}
\end{equation}

\begin{equation}
\varepsilon_{2}(k)=\frac{\varepsilon_{k}+\varepsilon_{k+Q}}{2}-\mu-\sqrt{(\frac{\varepsilon_{k}-\varepsilon_{k+Q}}{2})^2+|\Delta(k)|^2}
\end{equation}
 and
\begin{equation}
\Delta(k)=\frac{1}{N}\sum_{k^{\prime}\sigma}
V_{kk^\prime}\frac{\Delta(k^\prime)}{2E(k^\prime)}
(\tanh (\frac{\varepsilon_{1}(k^\prime)}{2T})-
\tanh (\frac{\varepsilon_{2}(k^\prime)}{2T}))
\end{equation}
 
   When $\mu$ crosses only the lower branch ($\varepsilon_{2}(k)$), the system can remain metallic and has pseudogap. The pseudogap $\Delta_{PS}(k_F)$ is given by

\begin{eqnarray} 
\Delta_{PS}(k_F)
&=&
-\varepsilon_{2}(k_F)
\nonumber\\ 
&=& 
\frac{1}{2}\mu-\frac{1}{2}\varepsilon_{k_{F}+Q}+\sqrt{\frac{1}{4}(\mu-\varepsilon_{k_{F}+Q})^2+|\Delta(k_F)|^2} 
\end{eqnarray}

where $k_{F}\in\varepsilon_{k_F}-\mu=0$, i.e. on the Fermi surface as shown in Fig.4. This is the so-called ``pseudogap''.   

\vspace{.5cm}
\noindent {\bf Results and Discussion} 
\vspace{.5cm}

By Eqs.(13), Eqs.(14), Eqs.(15) and Eqs.(16), we calculated the angle dependence ($\phi=-\arctan (\frac{k_{F_x}}{k_{F_y}})$ ) of the pseudogap which is indicated by Eqs.(16) in the underdoped regime ($\mu>0$) are shown in Fig.5. For both the d-wave separable potential and Hubbard interaction we note that in the proximity of the saddle point the Fermi surface are open gap and with decreased doping the gapped portion extend (In Fig.5 it are represented by the thick sections.), while the Fermi surface in the necklace regions are almost gapless ( Because in calculation we have neglected the effect of life of elementary excitation due to various scattering which is strongest for these regions, it is not equal to zero.). In Fig.6 we plot theoretical hole doping concentration dependence of the maximum value of the pseudogap for both the d-wave separable potential and Hubbard interaction. We note that with decreased doping the maximum value of the pseudogap increase. Such the behaviors are close to the experimentally found behavior above $T_{c}^{19-21}$.

When $t^{\prime}=0$, then Eqs.(16) is changed into following form

\begin{equation}
\Delta^{\prime}_{PS}(k_F)=\mu+\sqrt{\mu^{2}+|\Delta(k_F)|^{2}}
\end{equation}

For both Hubbard interaction
and the d-wave separable potential the entire Fermi surface are open gap.
These are not close to the experimentally found behavior above
$T_{c}^{19-21}$ and suggest that $t^{\prime}$
plays important rale in formation of pseudogap in the underdoped
cuprate superconductors.

Now we study that the effect of pseudogap on
the spectral function. By the definition of the spectral
function we can get the spectral function of the following form
for the system:
 
\begin{equation}
A(k,\omega)=
2[\mu_{k}^{2}
\frac{\Gamma}{(\omega-\varepsilon_{1}(k))^{2}+\Gamma^{2}}+\upsilon_{k}^{2}\frac{\Gamma}{(\omega-\varepsilon_{2}(k))^{2}+\Gamma^{2}}]
\end{equation}
The standard $\delta$-function
$\delta(\omega-\varepsilon_{1}(k))$ and
$\delta(\omega-\varepsilon_{2}(k))$
have been replaced by Lorentz function with width
$\Gamma$. $\Gamma$ is a crud representation of broadening
due to interaction of the quasiparticle. Fig.7 presents function
$A(k_{F},\omega)$
calculated for $k$ changing along the Fermi surface at $\Gamma=0.04t$.
we can see that the spectral function has the clear double peak gap-like
structure in the vicinity of ($0,\pi$),
while the single peak structure of quasiparticles
exists in the vicinity($\pi/2,\pi/2$).
    
For the density of states 
\begin{equation}
N(\omega)=\frac{1}{N}\sum_{k\sigma}A(k,\omega)
\end{equation}
which is presented in Fig.8. We can see that the suppression of the density of states at the low energy is essential.  

From above discussing we find that for the pseudogap formation not only needs that there is the SDW energy gap, but also there are the special dispersion (i.e. there are the necklace regions). In our calculations we did not include the superconductivity and superconducting fluctuations and thus the pseudogap is not connected with the superconductivity and superconducting fluctuations. As for the coexistence problem of the pseudogap and the superconductivity and above the superconducting transition temperature $T_{c}^{19-21}$  the pseudogap being temperature dependent are presently in progress and will be reported elsewhere.

  In order to test our theory, here we suggest a testing experiment that the material has above discussing the special dispersion and in metal state but has not superconducting phase.

  In summary, in this paper we provides a basis for interpretation of the pseudogap formation in the underdoped cuprate superconductors and suggests that the pseudogap formation might been due to the unique interior of the underdoped cuprate.

\begin{table}
\begin{tabular}{cc}
$\varepsilon_{k}$ & $\varepsilon_{k+Q}$\\
\tableline
      $A_+$ & $B_-$\\ 
      $A_-$ & $B_+$\\ 
      $A_{+}^{\prime}$ & $B_{+}^{\prime}$\\
\end{tabular}
\end{table}

\begin{figure}[htbp]
\begin{center}
\includegraphics[width=10cm,angle=0]{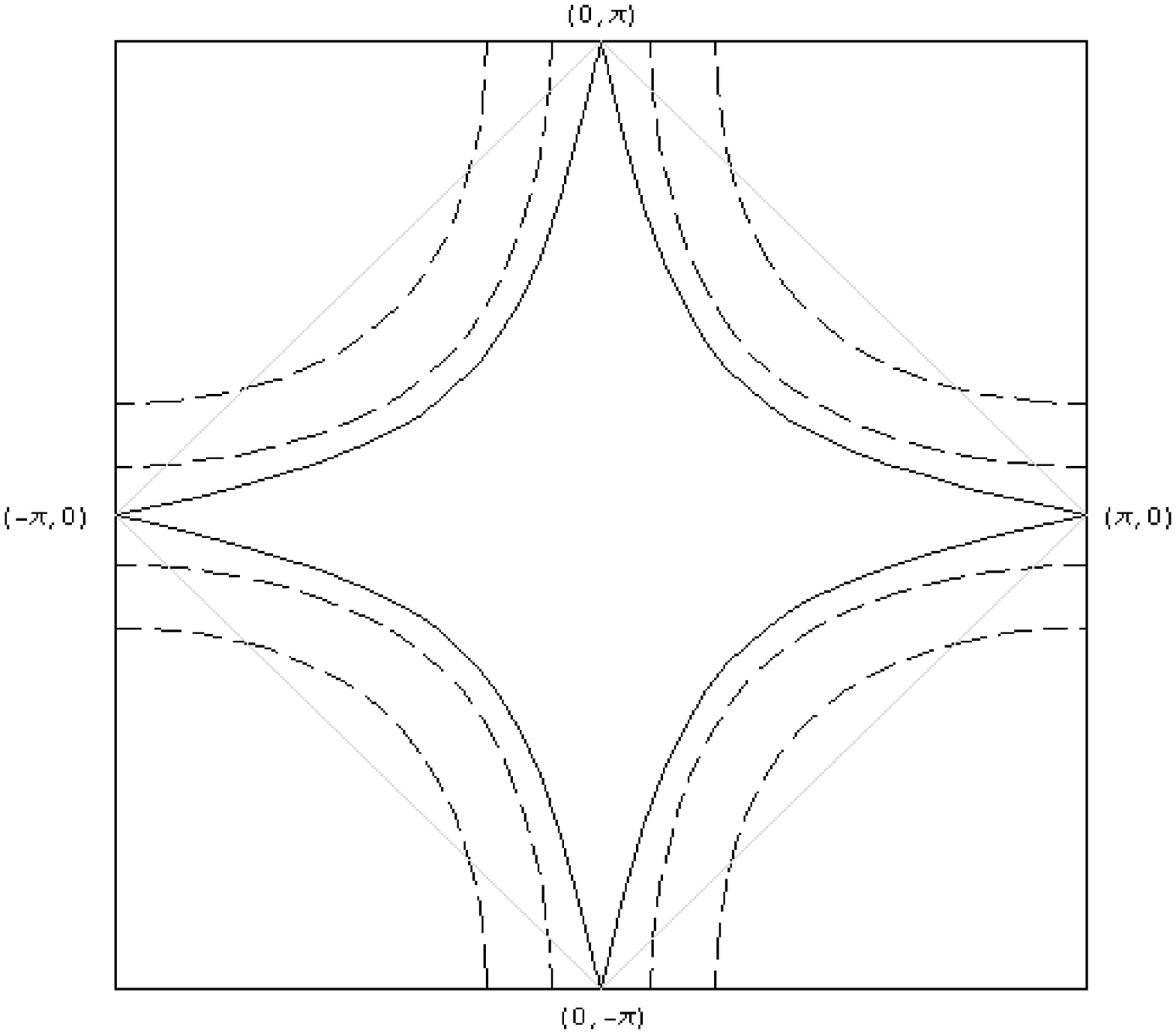}
\caption{ Fermi surface for the optimally doped (solid line) and underdoped (break lines) and. two different saddle points locate at ($\pm\pi,0$) and ($0,\pm\pi$). The square is the original Brillouin zone. The shadow line represents the magnetic Brillouin zone.}
\end{center}
\end{figure}

\begin{figure}[htbp]
\begin{center}
\includegraphics[width=10cm,angle=0]{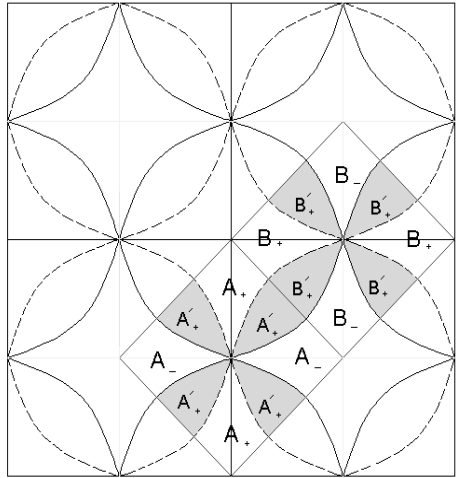}
\caption{The Brillouin zone: The square is the original Brillouin zone. The rectangle is our choice of first Brillouin zone, which consists of the region A and B including ($\pi,0$) and ($0,\pi$), respectively. The spectrum of the $\varepsilon_{k}=0$ and $\varepsilon_{k+Q}=0$  are represented by the solid curves and the break curves for $t^{\prime}/t=-0.4$, which relation is shown in Table-1. In regions $A_+$, $A^{\prime}_{+}$, $B_+$,  and $B^{\prime}_{+}$ $\varepsilon_{k}>0$ . In regions $A_{-}$ and $B_{-}$  $\varepsilon_{k}<0$. The shadow regions (i.e. $A^{\prime}_{+}$ and $B^{\prime}_{+}$) indicates the necklace regions (i.e. in these regions $\varepsilon_{k}>0$ , while $\varepsilon_{k+Q}>0$ ) of our choice of first Brillouin zone.
 }
\end{center}
\end{figure}

\begin{figure}[htbp]
\begin{center}
\includegraphics[width=10cm,angle=0]{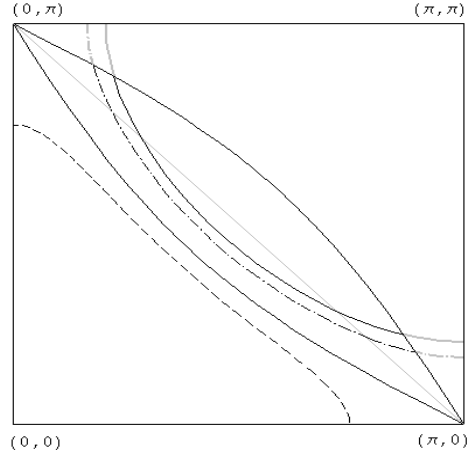}
\caption{The Fermi surfaces in the first quarter of the original Brillouin zone. The break cures represent the Fermi surface of the overdoped ($\mu<0$). The solid and dot-dashed cures represent the Fermi surface of the underdoped ($\mu>0$). The thick sections characterize the portions of the Fermi surface out of the necklance. }
\end{center}
\end{figure}

\begin{figure}[htbp]
\begin{center}
\includegraphics[width=10cm,angle=0]{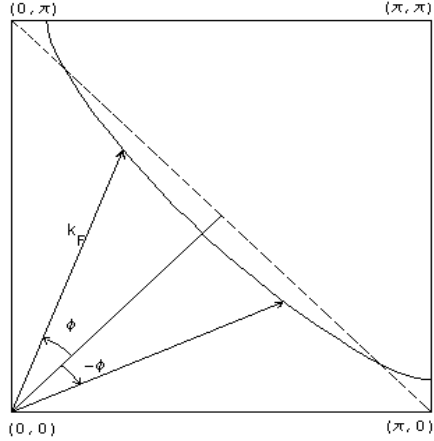}
\caption{$\pm\phi$ is angle determining the direction of electronic momentum $k$ on the Fremi surface and electronic Fermi momentum $k_F$. The solid curve represents the Fermi surface. The break line represents the first quarter of the original magnetic Brillouin zone boundary. The square is the first quarter of the original Brillouin zone.
 }
\end{center}
\end{figure}

\begin{figure}[htbp]
\begin{center}
\includegraphics[width=10cm,angle=0]{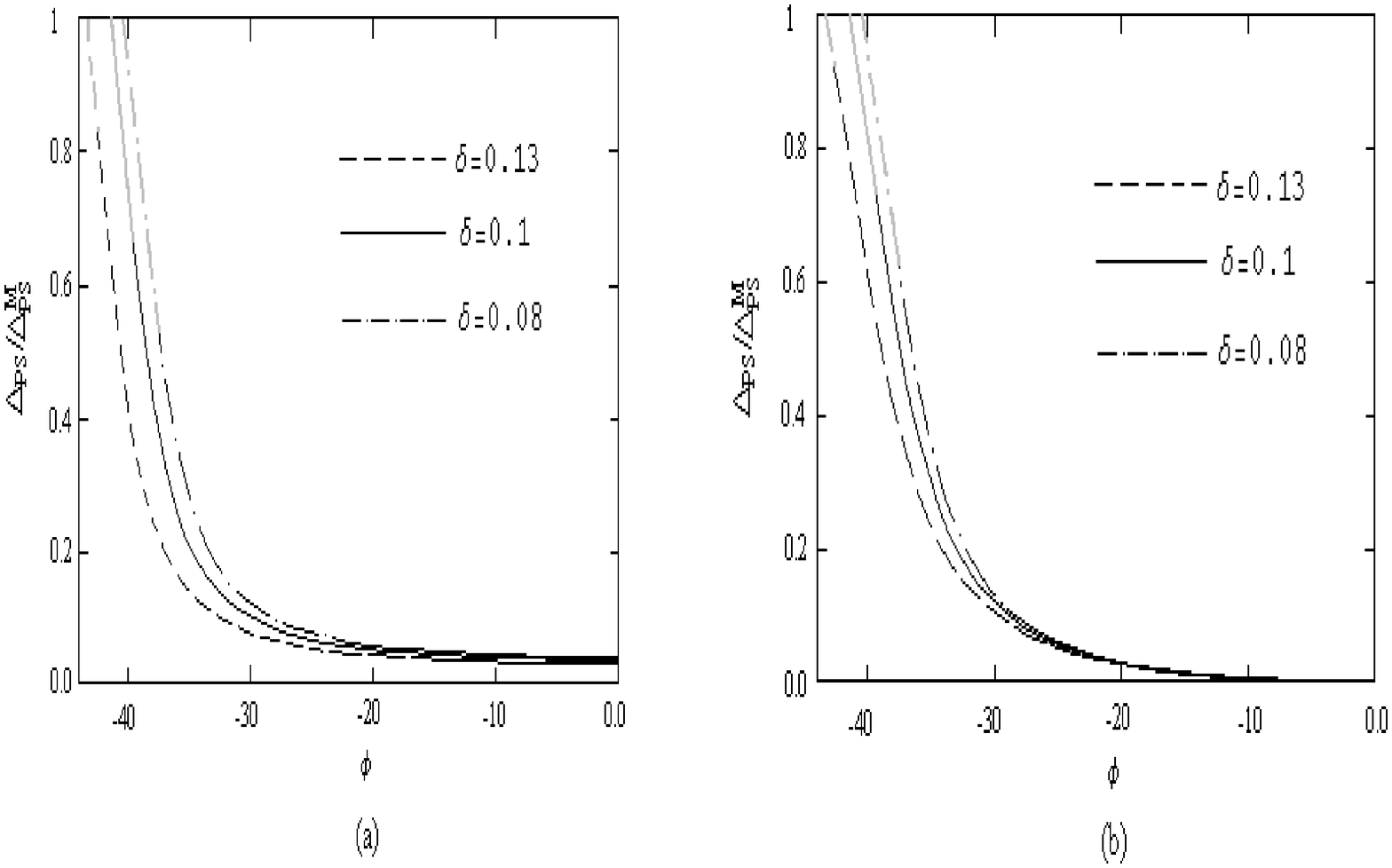}
\caption{Theoretical angle dependence of the pseudogap calculated by Eqs.(16) in the underdoped regime $\mu>0$ ($t^{\prime}/t=-0.16$), $\Delta^{M}_{PS}$ is the maximum value of the pseudogap. $\delta$ indicates the hole doping concentration. The thick sections represent the gap of out of the necklace regions: (a) for Hubbard interaction $U/t=1.6$; (b) for the d-wave separable potential $U/t=0.8$.
 }
\end{center}
\end{figure}

\begin{figure}[htbp]
\begin{center}
\includegraphics[width=10cm,angle=0]{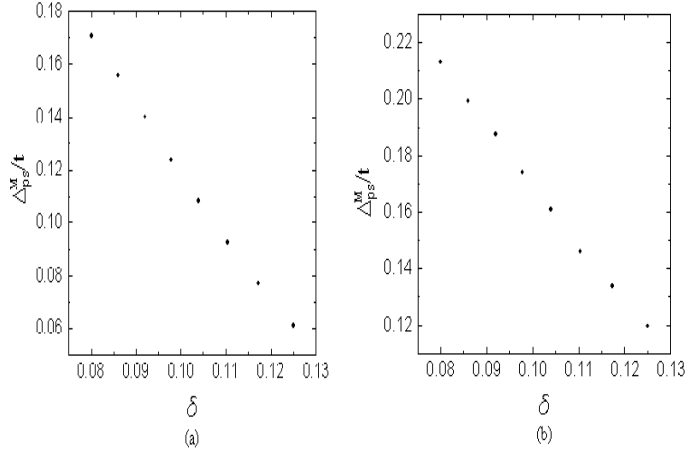}
\caption{Theoretical hole doping concentration dependence of the maximum value of the pseudogap calculated in the underdoped regime $\mu>0$ ($t^{\prime}/t=-0.16$), $\Delta^{M}_{PS}$  is the maximum value of the pseudogap. $\delta$  indicates the hole doping concentration: (a) for Hubbard interaction $U/t=1.6$;  (b) for the d-wave separable potential $U/t=0.8$.}
\end{center}
\end{figure}

\begin{figure}[htbp]
\begin{center}
\includegraphics[width=10cm,angle=0]{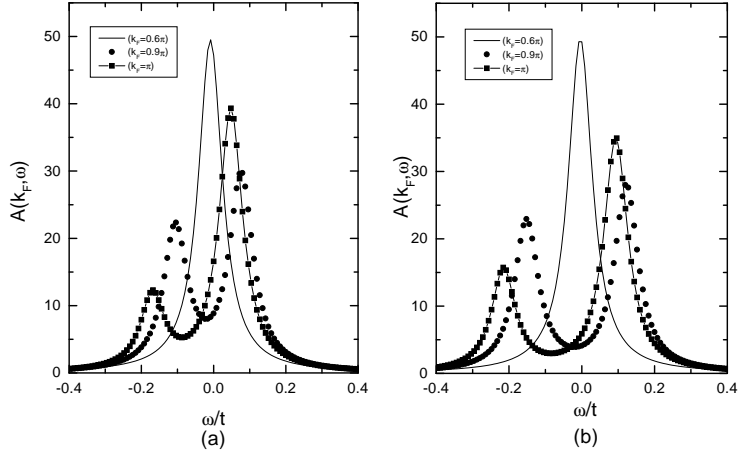}
\caption{The spectral function $A(k_{F},\omega)$ calculated for $k$ changing along the Fermi surface at $\Gamma=0.04t$ ($t^{\prime}/t=-0.16$, $\delta=0.08$): (a) for Hubbard interaction $U/t=1.6$;  (b) for the d-wave separable potential $U/t=0.8$.}
\end{center}
\end{figure}

\begin{figure}[htbp]
\begin{center}
\includegraphics[width=10cm,angle=0]{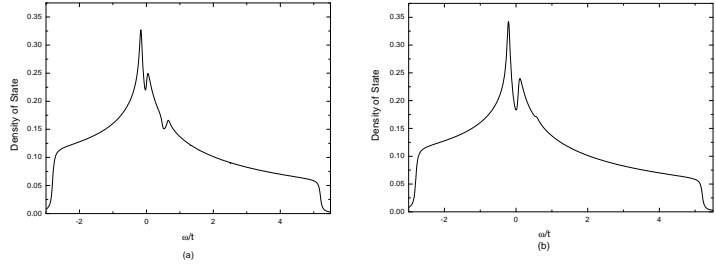}
\caption{The density of ststes $N(\omega)$ calculated by Eqs.(19) at $\Gamma=0.04t$ ($t^{\prime}/t=-0.16$, $\delta=0.08$): (a) for Hubbard interaction $U/t=1.6$;  (b) for the d-wave separable potential $U/t=0.8$.}
\end{center}
\end{figure}

\end{document}